\documentclass[%
 reprint,
 superscriptaddress,
 amsmath,amssymb,
 aps,
]{revtex4-1}

\usepackage{graphicx}
\usepackage{dcolumn}
\usepackage{bm}
\usepackage{xcolor}
\usepackage{url}
\usepackage{longtable}
\usepackage{cleveref}

\usepackage{subfigure}
\usepackage{multirow}

\begin{document}


\title{First Extraction of the $\phi$-$^{4}\mathrm{He}$ scattering length from near-threshold $\phi$ photoproduction on helium-4}


\author{Chengdong Han}
\email{chdhan@impcas.ac.cn}
\affiliation{Institute of Modern Physics, Chinese Academy of Sciences, Lanzhou 730000, China}
\affiliation{School of Nuclear Science and Technology, University of Chinese Academy of Sciences, Beijing 100049, China}
\affiliation{State Key Laboratory of Heavy Ion Science and Technology, Institute of Modern Physics, Chinese Academy of Sciences, Lanzhou 730000, China}

\author{Wei Kou}
\email{kouwei@impcas.ac.cn}
\affiliation{Institute of Modern Physics, Chinese Academy of Sciences, Lanzhou 730000, China}
\affiliation{School of Nuclear Science and Technology, University of Chinese Academy of Sciences, Beijing 100049, China}

\author{Rong Wang}
\email{rwang@impcas.ac.cn}
\affiliation{Institute of Modern Physics, Chinese Academy of Sciences, Lanzhou 730000, China}
\affiliation{School of Nuclear Science and Technology, University of Chinese Academy of Sciences, Beijing 100049, China}
\affiliation{State Key Laboratory of Heavy Ion Science and Technology, Institute of Modern Physics, Chinese Academy of Sciences, Lanzhou 730000, China}

\author{Xurong Chen}
\email{xchen@impcas.ac.cn}
\affiliation{Institute of Modern Physics, Chinese Academy of Sciences, Lanzhou 730000, China}
\affiliation{School of Nuclear Science and Technology, University of Chinese Academy of Sciences, Beijing 100049, China}
\affiliation{State Key Laboratory of Heavy Ion Science and Technology, Institute of Modern Physics, Chinese Academy of Sciences, Lanzhou 730000, China}
\affiliation{Southern Center for Nuclear Science Theory, Institute of Modern Physics,Chinese Academy of Sciences, Huizhou 516000, China}

\date{\today}

\begin{abstract}
  We present a model-independent extraction of the $\phi$–$^{4}\mathrm{He}$ scattering length from near-threshold $\phi$ photoproduction on helium-4,
  based on LEPS Collaboration data for the coherent process $\gamma + {}^{4}\mathrm{He} \to \phi + {}^{4}\mathrm{He}$ and the Vector Meson Dominance framework.
  Assuming an energy-independent differential cross section, we extract the absolute value of the $\phi-^4{\rm He}$ scattering length
  $|\alpha_{\phi^{4}\mathrm{He}}| = (3.33 \pm 0.06) \times 10^{-4}~\mathrm{fm}$ from a fit at threshold $t_{\text{thr}}$.
  This value is orders of magnitude smaller than those for $\phi N$ and $\phi d$ scattering length, indicating an inverse dependence of $|\alpha_{VA}|$ on the target nucleus mass.
  Our results provide new insight into the $\phi$-nucleus interaction, supporting the notion of weak $\phi$-nucleus coupling.
  We further explore the dependence of $|\alpha_{VA}|$ on the vector meson mass, the target atom mass, and the threshold energy.
  An approximate exponential suppression of $|\alpha_{VA}|$ is observed with increasing vector meson mass or target atom mass,
  indicating that heavier vector mesons or heavier target nuclei exhibit weaker couplings in vector meson–nucleus interactions.
\end{abstract}

\pacs{12.38.?t, 14.20.Dh}
\maketitle


\section{Introduction}
\label{introduction}
Understanding the interactions between vector meson and nucleon, as well as their extension to nuclear systems,
remains a fundamental challenge in non-perturbative Quantum Chromodynamics (QCD).
Since the discovery of the vector mesons, such as $\omega$, $\rho$,  $\phi$, $J/\psi$, $\psi(2S)$ and $\Upsilon$, they have served as powerful probes for the study of hadronic interaction.
In particular, the behavior of near-threshold photoproduction cross sections is closely related to the vector meson–nucleon scattering length~\cite{Gell-Mann:1961jim}. 
Experimentally, such interactions can be accessed through vector meson photoproduction within the framework of the Vector Meson Dominance (VMD) model \cite{Bauer:1977iq}. 
According to the VMD model, the absolute value of the scattering length can be can be extracted either from
the total near-threshold vector meson photoproduction cross-section or from the differential cross-section of vector meson photoproduction at threshold~\cite{Titov:2007xb}.
The VMD model has been successfully employed in describing near-threshold photoproduction processes, providing a quantitative connection between experimentally
measurable cross sections and the underlying vector meson–nucleon scattering lengths. Moreover, recent developments have demonstrated that these measurements
are also sensitive to the mass radius of nucleons and light nuclei~\cite{Wang:2021ujy, Kharzeev:2021qkd, Wang:2021dis, Han:2022qet, Wang:2023uek, Kou:2021bez}.
Scattering lengths for several vector mesons, including $\omega p$, $\phi p$, $J/\psi-p$, $\psi(2S) p$ and $\rho^{0} p$, have been extensively investigated
in Refs.\cite{Strakovsky:2014wja,Strakovsky:2019bev,Strakovsky:2020uqs,Pentchev:2020kao,Wang:2022xpw,Wang:2022zwz, Han:2022khg, Han:2022btd, Abreu:2024qqo}.

In addition, vector meson–deuteron interactions have also been studied using $\phi$, $\omega$, and $\rho^0$ meson photoproduction data,
with the extraction of $|\alpha_{\phi d}|$, $|\alpha_{\omega d}|$, and $|\alpha_{\rho^0 d}|$\cite{Wang:2022tuc, Wang:2023wlq}.
Predictions for the scattering lengths of heavier vector mesons such as $J/\psi$–$d$ and $\Upsilon$–$d$ have also been made~\cite{Wang:2023wlq}.

In this context, we report the first extraction of the $\phi$–$^4$He scattering length based on near-threshold $\phi$-meson photoproduction on helium-4. 
The analysis is performed using the currently available experimental data from the LEPS Collaboration \cite{LEPS:2017nqz}.
\textcolor{black}{For the vector meson photoproduction from helium-4 near threshold data, only the measurements from LEPS Collaboration are currently available, which results in limited statistical precision.}
Following the approach in Ref.\cite{Strakovsky:2019bev}, where the $J/\psi$–$p$ scattering length was estimated using the VMD model and GlueX data near-threshold $t_{thr}$,
we apply the same formalism here.
Within the framework of the VMD model, the total cross section for $\gamma N \rightarrow VN$ at threshold is related to the total $VN \rightarrow VN$ cross section
and the corresponding scattering length $|\alpha_{VN}|$, as given in Ref.\cite{Titov:2007xb}:
\begin{equation}
	\sigma^{\gamma N}\left(s_{t h r}\right)=\frac{\alpha \pi}{\gamma_{V}^{2}} \frac{q_{VN}}{k_{\gamma N}} \cdot \sigma^{VN}\left(s_{t h r}\right)=\frac{\alpha \pi}{\gamma_{V}^{2}} \frac{q_{VN}}{k_{\gamma N}} \cdot 4 \pi \alpha_{VN}^{2},
	\label{eq:total}
\end{equation}
where $\alpha$ = 1/137 is the fine-structure constant,
and $V$ denotes the vector meson species (e.g., $\omega$, $\rho$, $\phi$, $J/\psi$, etc). 
The $k_{\gamma N}$ in the above equation is the initial momentum in the center-of-mass frame,
\begin{equation}
  k_{\gamma N} = \frac{1}{2W} \sqrt{W^4 - 2(m_{\gamma}^2 + m_N^2)W^2 + (m_{\gamma}^2 - m_N^2)^2},
\end{equation}
and $q_{VN}$ in the above equation is the final momentum in the center-of-mass frame,
\begin{equation}
  q_{VN} = \frac{1}{2W} \sqrt{W^4 - 2(m_{V}^2 + m_N^2)W^2 + (m_{V}^2 - m_N^2)^2},
\end{equation}
with the $W$ being the total invariant mass of the real photon and target atom collision system, defined as $W=\sqrt{m_N^2+2m_N E_{\gamma}}$, and the $E_{\gamma}$ incident photon energy.
The photon–vector meson coupling constant, $\gamma_V$, is determined from the decay width of $V \rightarrow e^+e^-$.
The photon–$\phi$ coupling constant used in this analysis is $\gamma_\phi = 6.71$~\cite{Workman:2022ynf}.
Eq.(\ref{eq:total}) is evaluated at the threshold energy, where $s_{thr}$ = $(m_{V} + m_{N})^{2}$ with $m_{V}$ and $m_{N}$ being the masses of the vector meson and the target atom,
respectively. 

In the study, we focus on estimating the scattering length of $\phi$ meson interacting with the helium-4. 
To this end, we utilize the experimental measurements of the differential photoproduction cross-section of $\phi$-$^{4}\mathrm{He}$.
Specifically, we adopt the theoretical relation between the total and differential cross sections at threshold, as outlined in Ref.~\cite{Pentchev:2020kao}.
The connection between the differential cross section and the scattering length is given by:
\begin{equation}
  \begin{split}
    \frac{d \sigma^{\gamma N}}{d t}\left(s_{t h r}, t=t_{thr}\right)=\frac{\alpha \pi}{\gamma_{V}^{2}} \frac{\pi}{k_{\gamma N}^{2}} \cdot \alpha_{ VN}^{2}(t=t_{thr}), \\
    \frac{d \sigma^{\gamma N}}{d t}\left(s_{t h r}, t=0\right)=\frac{\alpha \pi}{\gamma_{V}^{2}} \frac{\pi}{k_{\gamma N}^{2}} \cdot \alpha_{ VN}^{2}(t=0),
    \label{eq:diffxsection}
  \end{split}
\end{equation}
where the $t_{thr}$ is the production threshold of vector meson, defined as $t_{thr}$=$-m_{V}^2 m_{N}/(m_V+m_N)$.
A major challenge in determining the scattering length at threshold is the need to extrapolate the cross section to either $t \to t_{\text{thr}}$ or $s \to s_{\text{thr}}$,
since the differential cross section $d\sigma^{\gamma N}/dt$ on the left-hand side of Eq.~(\ref{eq:diffxsection}) is not directly measurable.
This requires either extrapolating the energy to the production threshold or extending the $t$ variable from the physical region ($t_{\text{min}} < t < t_{\text{max}}$)
to the unphysical point $t = 0$.
In the present work, the vector meson–nucleon scattering length is extracted from the differential cross-section data by extrapolating only to the energy at the production threshold, $t_{\text{thr}}$.

In this work, the differential cross-section data for near-threshold $\phi$ photoproduction on helium-4 target were fitted using the following exponential function,
\begin{equation}
\begin{split}
\frac{d\sigma}{dt}=Ae^{-bt},
\end{split}
\label{eq:exp_fit}
\end{equation}
where $A=d\sigma/dt|_{t=0}$ represents the forward differential cross-section, and $b$ is the slope parameter.
By combining Eq. (\ref{eq:diffxsection}) and Eq. (\ref{eq:exp_fit}),
the value of the forward differential cross-section at the production threshold, $d\sigma/dt|_{t=t_{thr}}$, can be obtained,
from which the $\phi$-$^{4}\mathrm{He}$ scattering length $|\alpha_{\phi^{4}\mathrm{He}}|$ is extracted.

Significant progress has been achieved in determining the scattering lengths of vector meson with proton, neutron, and deuteron.
However, experimental and theoretical knowledge of vector meson–$^4$He scattering lengths remains limited.
In previous analyses, the scattering lengths of vector meson-nucleon and vector meson-deuteron have been extracted from near-threshold photoproduction data
of $\omega$, $\phi$, $J/\psi$, $\psi(2S)$, and $\rho^0$ mesons \cite{Ishikawa:2019rvz,Titov:2007xb,Strakovsky:2019bev,Pentchev:2020kao,Wang:2022xpw,Wang:2022zwz,
  Han:2022khg,Han:2022btd,Wang:2022tuc,Wang:2023wlq}.
Motivated by these developments, we attempt to extract the vector meson–$^4$He scattering length using near-threshold photoproduction data on a helium-4 target,
in order to investigate the correlation between the scattering length and both the mass of the target atom and the production threshold energy $\sqrt{s_{thr}}$.

\section{$\phi-^4{\rm He}$ scattering length from $\phi$ differential photoproduction cross sections}
\label{data_analysis_scattering_lengths}

\begin{figure}[htp]
\centering
\includegraphics[width=0.50\textwidth]{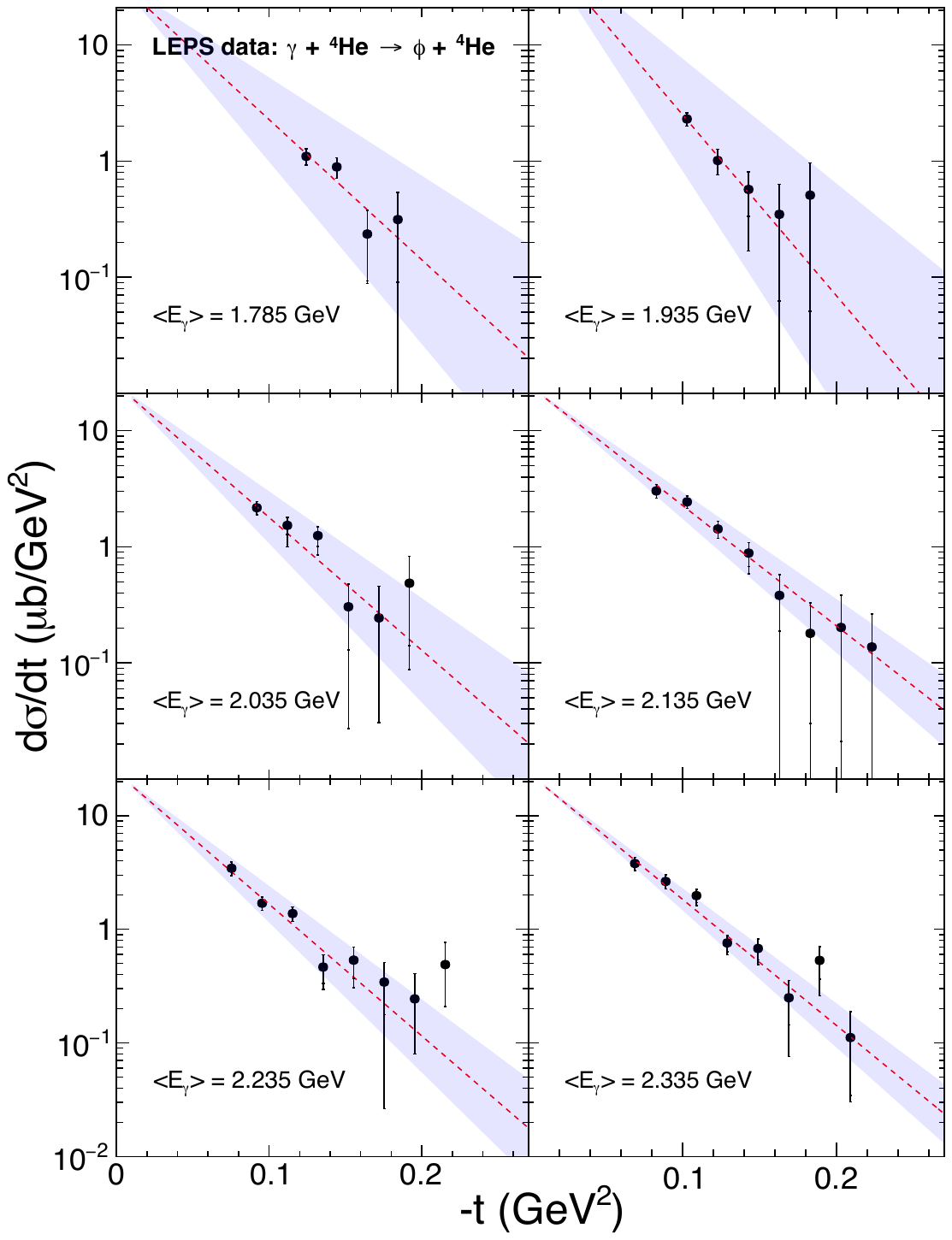}
\caption{
  Differential cross sections of the coherent $\phi$ meson photoproduction ($\gamma$ + $^{4}$He $\rightarrow$ $\phi$ + $^{4}$He)
  near the threshold as a function of the momentum transfer $-t$ off the helium-4 target \cite{LEPS:2017nqz}.
  The six incident photon energies $E_{\gamma}$ (1.785, 1.935, 2.035, 2.135, 2.235 and 2.335 GeV) near the threshold of incoherent $\phi$ meson photoproduction are marked in the figure.
  \textcolor{black}{The smaller error bars on the vertical axis indicate the statistical error, while the larger ones correspond to the sum of the statistical and systematic errors in quadrature.}
}
\label{fig:Phi_nucleon_Rm}
\end{figure}

\begin{table*}
  \caption{
    The absolute value of the extracted $\phi-^4{\rm He}$ scattering length obtained from
    the differential cross sections of coherent $\phi$ mesons photoproduction from the helium-4
    near threshold at different photon energies.
  }
\begin{center}
  \begin{ruledtabular}
    \begin{tabular}{c ccc}
      $E_{\gamma}$ (GeV) & 1.785 & 1.935 & 2.035 \\
      $|\alpha_{\phi ^{4}\mathrm{He}}|$ (fm)
      & $2.28\times10^{-3}\pm7.88\times10^{-5}$
      & $1.39\times10^{-4}\pm6.39\times10^{-6}$
      & $3.76\times10^{-3}\pm7.99\times10^{-5}$ \\
      \hline
      $E_{\gamma}$ (GeV) & 2.135 & 2.235 & 2.335 \\
      $|\alpha_{\phi ^{4}\mathrm{He}}|$ (fm)
      & $1.05\times10^{-2}\pm1.16\times10^{-4}$
      & $3.42\times10^{-3}\pm5.21\times10^{-5}$
      & $5.54\times10^{-3}\pm5.11\times10^{-5}$ \\
    \end{tabular}
  \end{ruledtabular}
\end{center}
\label{tab:RadiusListPhiMeson}
\end{table*}

Figure~\ref{fig:Phi_nucleon_Rm} presents the differential cross sections for \textcolor{black}{coherent} $\phi$-meson photoproduction from helium-4 as a function of the momentum transfer $-t$.
The differential cross section data for coherent $\phi$-meson photoproduction were collected by the LEPS spectrometer using incident photons in the
energy range $E_\gamma = 1.685$--$2.385$~GeV~\cite{LEPS:2017nqz}.
The observed $|t|$-dependent differential cross sections are fitted using an exponential form.
Specifically, the parameters $A$ and $b$ were extracted by fitting Eq.(\ref{eq:exp_fit}) to the coherent $\phi$-meson photoproduction data
on the $^4$He target at six photon energies $E_{\gamma}$ (1.785, 1.935, 2.035, 2.135, 2.235 and 2.335 GeV)~\cite{LEPS:2017nqz}. 
Using the extracted fit parameters $A$ and $b$, we evaluated the forward differential cross section at threshold, $d\sigma^{\gamma ^{4}\mathrm{He}}/dt(s_{thr},t_{thr})$,
and subsequently determined the scattering length $|\alpha_{\phi\,^{4}\mathrm{He}}|$ at threshold via Eq.~(\ref{eq:exp_fit}).
The extracted scattering length $|\alpha_{\phi ^{4}\mathrm{He}}|$ with six incident photon energies are listed in Table \ref{tab:RadiusListPhiMeson}.
The weighted average of the scattering length for coherent $\phi$ meson photoproduction on $^{4}$He at different $E_{\gamma}$ energies
at $t=t_{thr}$ is calculated to be $|\alpha_{\phi\,^{4}\mathrm{He}}|$ = $(3.33 \pm 0.06) \times 10^{-4}\,\mathrm{fm}$.
\textcolor{black}{The weighted average was calculated using the standard error-weighted formula: $\overline{x}$ $\pm$ $\delta$ $\overline{x}$ = $\Sigma_{i}$$\omega_{i}$$x_{i}$/$\Sigma_{i}$$\omega_{i}$$\pm$
($\Sigma_{i}$$\omega_{i}$)$^{-1/2}$ with $\omega_{i}$ = 1/($\delta x_{i}$)$^{2}$ \cite{Workman:2022ynf}.
Notably, the weighted average value is consistent with the result obtained from a simultaneous fit to all differential cross-section data sets.}

\begin{figure}[htp]
\centering
\includegraphics[width=0.5\textwidth]{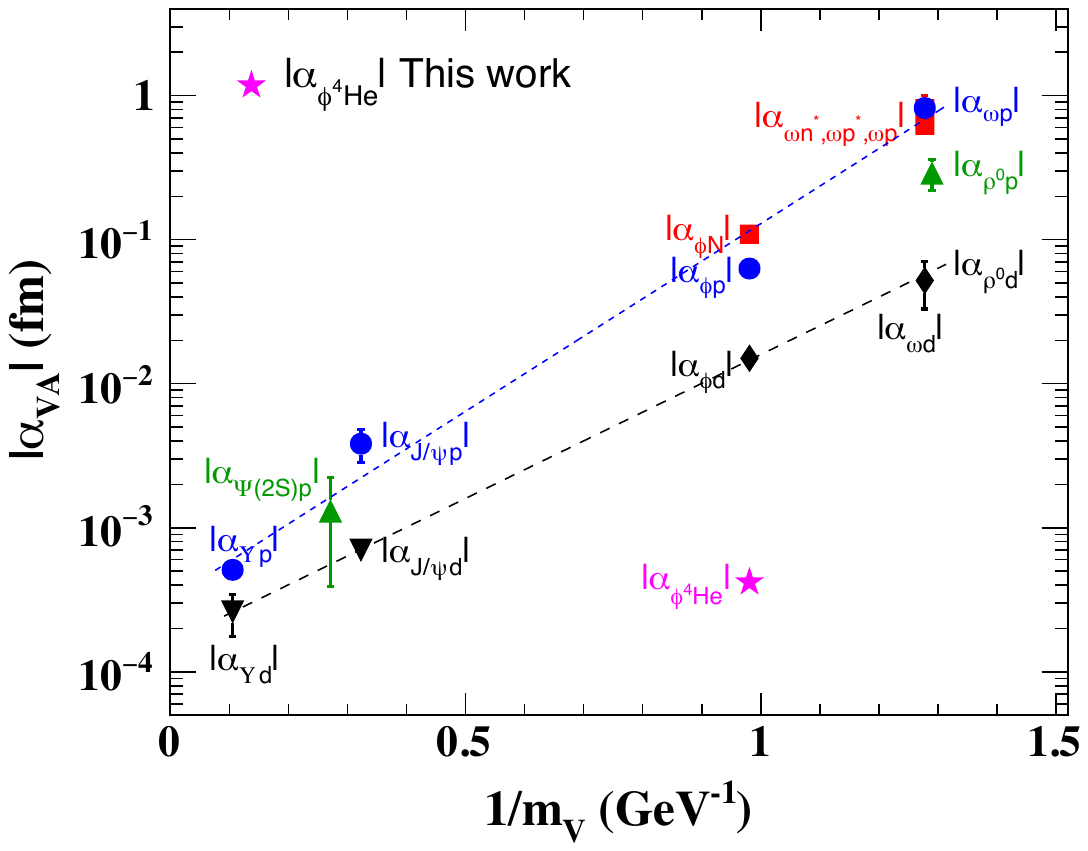}
\caption{
  The scattering length of $\phi$-$^{4}$He interation obtained from the differential cross section of $\phi$ photoproduction off a helium-4 based on the VMD model,
  as shown by the magenta star marker.
  Comparison of the scattering lengths $|\alpha_{VA}|$ as a function of the inverse mass of vector mesons,
  including $\omega$, $\rho^{0}$, $\phi$, $J/\psi$, $\psi(2S)$ and $\Upsilon$.
  The red squares show the $|\alpha_{\omega n^{*}}|$, $|\alpha_{\omega p^{*}}|$, $|\alpha_{\omega p}|$ and $|\alpha_{\phi N}|$ scattering length at threshold $t_{thr}$
  from our previous work \cite{Han:2022khg}.
  The blue cricles show the Strakovsky's analysis of $\omega$ \cite{Strakovsky:2014wja}, $\phi$ \cite{Strakovsky:2020uqs}, $J/\psi$ \cite{Pentchev:2020kao} and $\Upsilon$ \cite{Strakovsky:2021vyk}.
  The green triangles shows the Wang's analysis $\rho^{0}$ \cite{Wang:2022zwz} and $\psi(2S)$ \cite{Wang:2022xpw}.
  The long blue dashed line is hypothetical for vector meson-nucleon scattering length.
  The black diamonds for $|\alpha_{\phi d}|$, $|\alpha_{\omega d}|$ and $|\alpha_{\rho^{0} d}|$ are taken from Wang's work \cite{Wang:2022tuc, Wang:2023wlq};
  The black up triangles for $|\alpha_{\Upsilon d}|$ and $|\alpha_{J/\psi d}|$ are taken from Wang's prediction \cite{Wang:2023wlq};
  The long black dashed line is hypothetical for vector meson-deuteron scattering length.
}
\label{fig:avN_1mv}
\end{figure}

\begin{figure}[htp]
\centering
\includegraphics[width=0.5\textwidth]{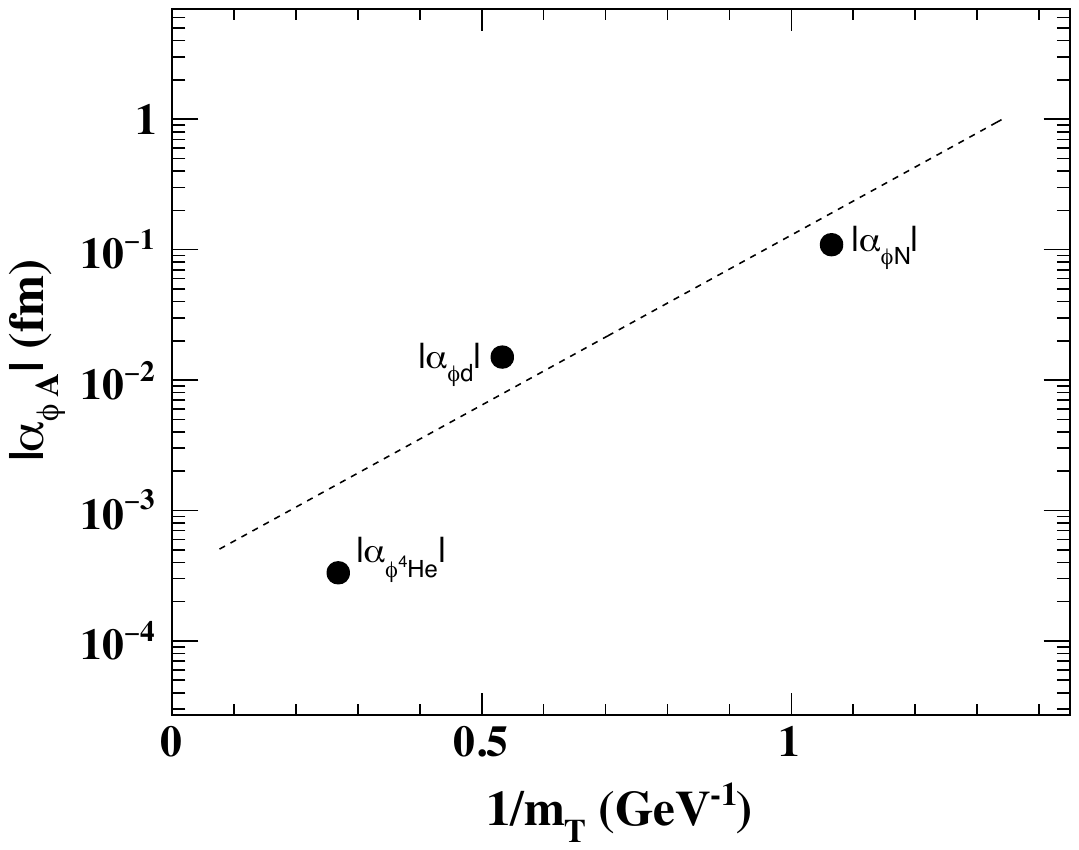}
\caption{
  The scattering lengths of $\phi$-N, $\phi$-d and $\phi$-$^{4}$He interation as function as the the inverse of the target atom mass.
  Comparison of the scattering lengths $|\alpha_{VN}|$ as a function of the inverse mass of vector mesons,
  including $|\alpha_{\phi N}|$ scattering length at threshold $t_{thr}$ from our previous work \cite{Han:2022khg}, $|\alpha_{\phi d}|$ from Wang's work \cite{Wang:2022tuc},
  and $\phi$-$^{4}$He from our this work.
  The long black dashed line is hypothetical for vector meson-nuleon and vector meson-nucleus scattering length,
  with the empirical formula being $|\alpha_{\phi N}| \sim 10^{-A}$, where $A$ is the atomic mass number treated as a free parameter.
}
\label{fig:aphiN_1mT}
\end{figure}

\begin{figure}[htp]
\centering
\includegraphics[width=0.5\textwidth]{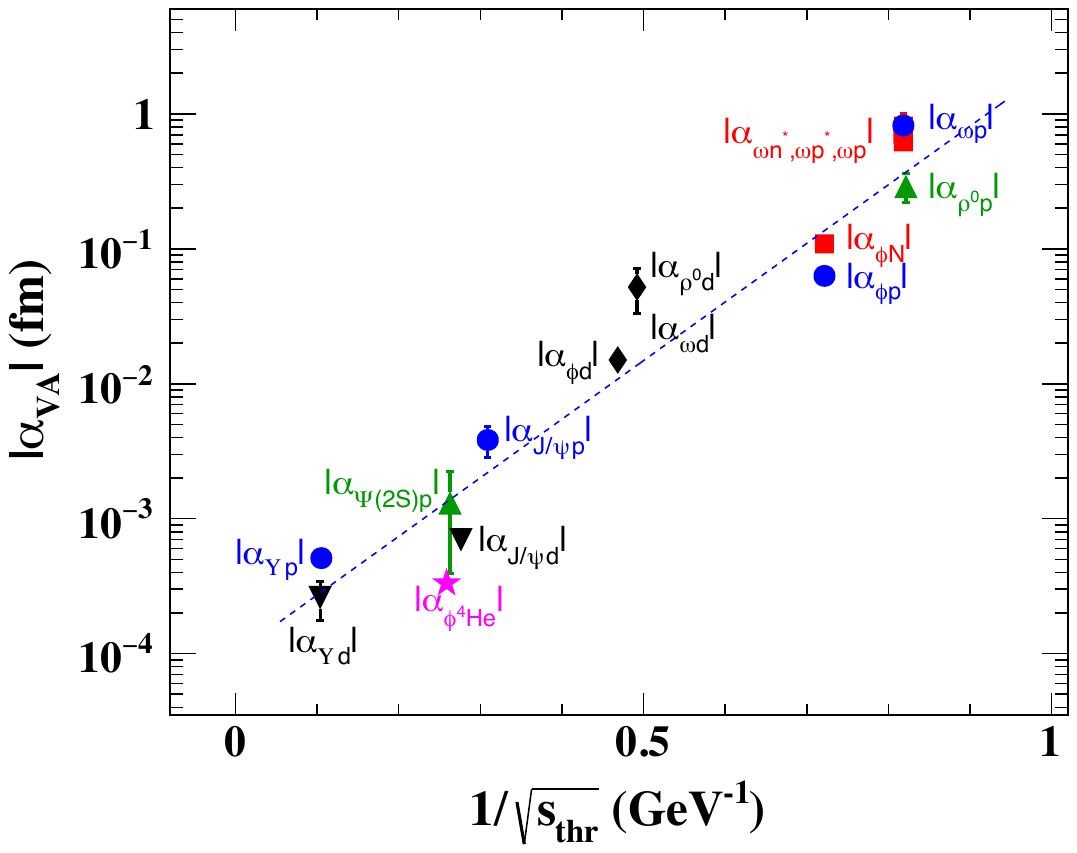}
\caption{
  The scattering lengths of vector meson-nuleon, vector meson-deuteron and vector meson-$^{4}$He interation as function as the the inverse of the threshold energy $\sqrt{s_{thr}}$,
  where $s_{thr}$ = $(m_{V} + m_{N})^{2}$ with $m_{V}$ and $m_{N}$ being the masses of the vector meson and the target atom, respectively.
  The long blue dashed line is hypothetical for vector meson-nuleon and vector meson-nucleus scattering length, with the assumed formula
  being $|\alpha_{VA}|$ $\sim$ $p_{0} \times e^{\frac{p_1}{\sqrt{s_{thr}}}}$, where $p_0$ and $p_1$ are assumed free parameters.
}
\label{fig:avN_1sqrtSthr}
\end{figure}

\begin{figure}[htp]
\centering
\includegraphics[width=0.5\textwidth]{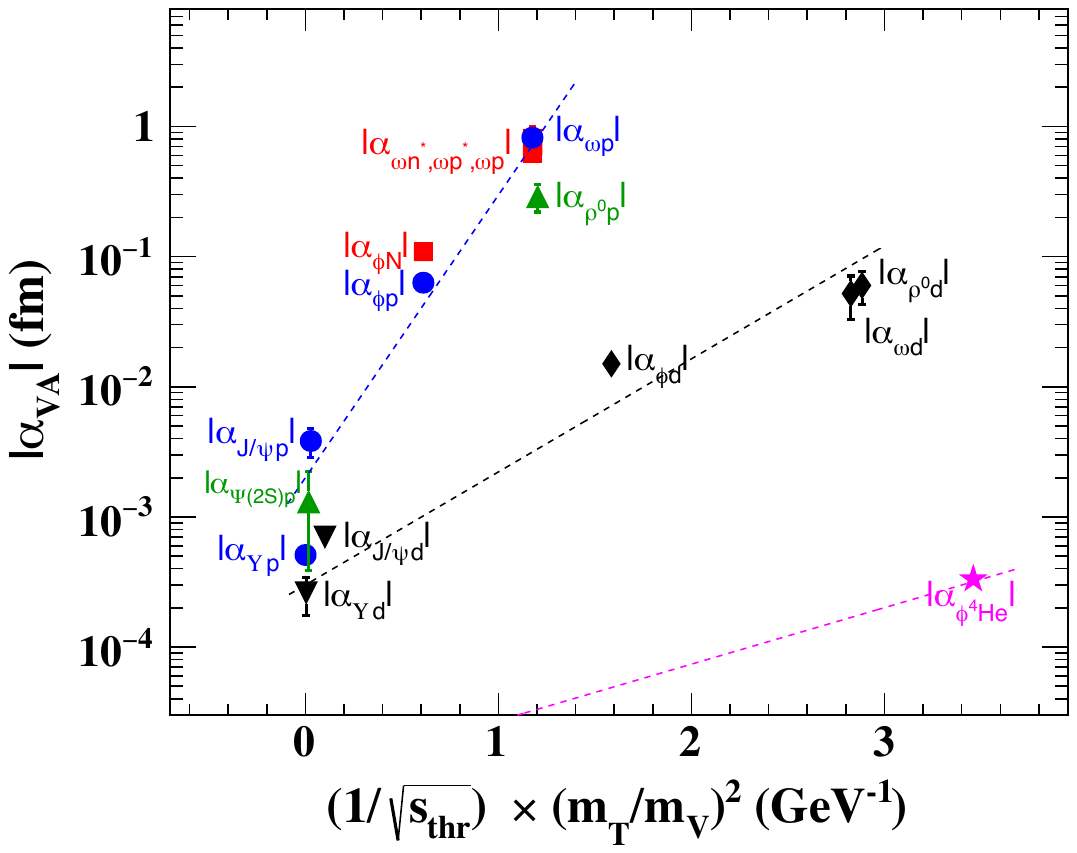}
\caption{
  The scattering lengths of vector meson-nuleon, vector meson-deuteron and vector meson-$^{4}$He interation as function as the the inverse of the product of the the threshold energy $\sqrt{s_{thr}}$
  and the square of the ratio of target atom mass $m_{T}$ to vector meson mass $m_{V}$.
  These long dashed lines are hypothetical for vector meson-nucleon and vector meson-nucleus scattering length,
  with the assumed formula being $|\alpha_{VA}|$ $\sim$ $p_0 \times e^{\frac{p_1}{\sqrt{s_{thr}}}\times (\frac{m_{T}}{m_{V}})^{2}}$, where $p_0$ and $p_1$ are assumed free parameters.
  Specifically, the blue dashed line represents the assumption for the vector meson-nucleon
  scattering length,
  the black dashed line represents the assumption for the vector meson-deuteron
  scattering length,
  the magenta dashed line represents the assumption for the vector meson-$^{4}$He
  scattering length. 
}
\label{fig:avN_1sqrtSthr_mN2_mV2}
\end{figure}

\section{Discussion}
\label{Discussion}
In order to investigate the correlation between the vector meson–nucleon (or nucleus) scattering length $|\alpha_{VN}|$ and the mass of the vector meson,
we compile in Fig.\ref{fig:avN_1mv} a collection of previously extracted scattering lengths
for various vector mesons ($\omega$, $\rho^{0}$, $\phi$, $J/\psi$, $\psi(2S)$, and $\Upsilon$)
interacting with different nuclear targets (proton and deuteron). These results are obtained using different extrapolation models and datasets,
as reported by Strakovsky\cite{Strakovsky:2014wja, Strakovsky:2020uqs, Pentchev:2020kao, Strakovsky:2021vyk} and Wang~\cite{Wang:2022zwz,Wang:2022xpw}.
Additionally, our previously extracted results for the scattering lengths $|\alpha_{\omega n^{}}|$, $|\alpha_{\omega p^{}}|$, $|\alpha_{\omega p}|$,
and $|\alpha_{\phi N}|$ are also included for comparison.
Fig.~\ref{fig:avN_1mv} illustrates the dependence of $|\alpha_{VA}|$ on the inverse mass of the corresponding vector meson.
An approximate exponential decrease of the scattering length with increasing meson mass is observed.
Specifically, heavier vector mesons exhibit weaker interaction strengths with nucleon or nuclei, as reflected by smaller values of $|\alpha_{VA}|$.
For the proton target, the scattering lengths follow the ordering: $|\alpha_{\Upsilon p}|$<$|\alpha_{\psi(2S) p}|$<$|\alpha_{J/\psi p}|$<$|\alpha_{\phi p}|$<$|\alpha_{\omega p}|$;
Similarly, for the deuterium target, the trend is: $|\alpha_{\Upsilon d}|$<$|\alpha_{J/\psi d}|$<$|\alpha_{\phi d}|$<$|\alpha_{\omega d}|$<$|\alpha_{\rho^{0} d}|$.
These patterns highlight a clear inverse correlation between the mass of the vector meson and its corresponding scattering length with a given nuclear target.
\textcolor{black}{The scaling laws for the vector meson-nucleon and vector meson-deuteron scattering lengths, as depicted by the blue and black dashed lines in this figure, respectively,
are both empirical trends.}

Figure~\ref{fig:aphiN_1mT} presents the scattering lengths for $\phi$--N, $\phi$--d, and $\phi$--$^{4}$He interactions as a function of the inverse of the target atom mass.
The comparison includes the scattering lengths $|\alpha_{VA}|$ as a function of the inverse mass of the target nucleus ($1/m_{\mathrm{T}}$),
incorporating $|\alpha_{\phi N}|$ at threshold momentum transfer $t_{\rm thr}$ from our previous work~\cite{Han:2022khg},
$|\alpha_{\phi d}|$ from Wang's study~\cite{Wang:2022tuc}, and the newly extracted $|\alpha_{\phi ^{4}{\rm He}}|$ from the present analysis.
\textcolor{black}{The long black dashed line represents a hypothetical trend for the vector meson–nucleon and vector meson–nucleus scattering lengths.}
As evident from the figure, the scattering length $|\alpha_{\phi A}|$ decreases with increasing target atom mass,
indicating that the $\phi$ meson interacts more weakly with heavier nuclear targets.
A clear hierarchy is observed among the scattering lengths:
$|\alpha_{\phi ^{4}{\rm He}}| < |\alpha_{\phi d}| < |\alpha_{\phi N}|$.
This trend, from Fig.~\ref{fig:avN_1mv} and \ref{fig:aphiN_1mT}, suggests that the scattering length reflects the effective interaction
between the vector meson probe and the target nucleus. Both the increasing mass of the vector meson and that of the target nucleus
tend to reduce the scattering length, highlighting a mass-dependent suppression of the interaction strength.
Furthermore, the dependence of the $\phi$--nucleus scattering length on the atomic mass number $A$ approximately follows an exponential trend of the form:
$|\alpha_{\phi A}| \sim 10^{-A}$,
indicating a rapid decrease in scattering strength with increasing target atom size.

Figure~\ref{fig:avN_1sqrtSthr} shows the extracted scattering lengths for vector meson–nucleon, vector meson–deuteron, and vector meson–$^4$He interactions
as a function of the inverse threshold energy $\sqrt{s_{\mathrm{thr}}}$, where $s_{\mathrm{thr}} = (m_{V} + m_{N})^{2}$,
with $m_{V}$ and $m_{N}$ being the masses of the vector meson and the target atom, respectively.
Regardless of whether the vector meson or the target nucleus is light or heavy, the scattering length $|\alpha_{VA}|$ exhibits a clear correlation
with $\sqrt{s_{\mathrm{thr}}}$, showing a trend approximately proportional to this inverse threshold energy. 
\textcolor{black}{The blue dashed line in this figure represents a phenomenological extrapolation for the vector meson–nucleon and vector meson–nucleus scattering lengths, based on the empirical formula}
$|\alpha_{VA}|$ $\sim$ $ p_0 \times \exp\left(\frac{p_1}{\sqrt{s_{\mathrm{thr}}}}\right)$,
where the $p_0$ and $p_1$ are assumed free parameters, set to $p_0$ = $1.0 \times 10^{-4}$ and $p_1$ = 10.0. 
These results reveal an approximate exponential suppression of $|\alpha_{VA}|$ with increasing vector meson and target nuclear mass,
suggesting that vector meson–nucleus couplings become progressively weaker for heavier mesons (e.g., $J/\psi$, $\psi(2S)$, $\Upsilon$) and heavier nuclear targets (e.g., $^{4}$He).

Figure~\ref{fig:avN_1sqrtSthr_mN2_mV2} presents the extracted scattering lengths for vector meson–nucleon, vector meson–deuteron, and vector meson–$^4$He interactions
as a function of $\left( \frac{1}{\sqrt{s_{\mathrm{thr}}}} \right) \times \left( \frac{m_T}{m_V} \right)^2$,
which corresponds to the inverse of the product of the threshold energy $\sqrt{s_{\mathrm{thr}}}$ and the square of the mass ratio between the target nucleus ($m_T$)
and the vector meson ($m_V$).
The observed correlation in Fig.\ref{fig:avN_1sqrtSthr_mN2_mV2} is motivated by the dependence of the vector meson–nucleon scattering length $|\alpha_{VN}|$ on the
trace anomaly contribution to the nucleon mass, $T_A$, as discussed in Eq.(13) of Ref.~\cite{Han:2022btd}.
\textcolor{black}{The long dashed lines represent hypothetical parametrizations of the vector meson–nucleon and vector meson–nucleus scattering lengths, assuming an empirical functional form}
$|\alpha_{VA}|$ $\sim$ $p_0 \times e^{\frac{p_1}{\sqrt{s_{thr}}}\times (\frac{m_{T}}{m_{V}})^{2}}$, where $p_0$ and $p_1$ are assumed phenomenological parameters.
Specifically, the blue dashed line denotes the assumption for the vector meson–nucleon scattering length with $p_0 = 2.0 \times 10^{-3}$ and $p_1 = 5.0$;
the black dashed line corresponds to the vector meson–deuteron scattering length with $p_0 = 3.0 \times 10^{-4}$ and $p_1 = 2.0$;
and the magenta dashed line represents the vector meson–$^4$He scattering length with $p_0 = 1.0 \times 10^{-5}$ and $p_1 = 1.0$.
These results suggest an approximately exponential dependence of the vector meson–nucleus scattering length
on the variable $\left( \frac{1}{\sqrt{s_{\mathrm{thr}}}} \right) \times \left( \frac{m_T}{m_V} \right)^2$,
indicating a possible universal scaling behavior across different nuclear targets.

\section{Summary}
\label{Summary}
In this work, we investigated the $\phi$-$^{4}\mathrm{He}$ scattering length from near-threshold $\phi$-meson photoproduction on a helium-4 target within the framework of the VMD model,
assuming an energy-independent differential cross section. By fitting the LEPS differential cross-section data with an exponential parameterization and extrapolating
to the threshold momentum transfer $t_{\mathrm{thr}}$, we extracted the scattering length for the $\phi$--$^4\mathrm{He}$ interaction for the first time.
The extracted value of the scattering length, $|\alpha_{\phi-^4{\rm He}}| = (3.33 \pm 0.06) \times 10^{-4}~\mathrm{fm}$, reflects a weak $\phi$--nucleus interaction,
significantly smaller than that observed for lighter nuclear targets such as the nucleon or deuteron.
Our results offer new insights into the $\phi$–nucleus interaction, reinforcing the hypothesis of a weak coupling between the $\phi$ meson and the nucleus.
This supports the expected trend of a decreasing meson-nucleus scattering length
with the increasing of target atom mass, which can be attributed to the spatial delocalization and reduced overlap of the meson wave function with heavier nuclear systems.
\textcolor{black}{However, assuming that the differential cross section near the threshold is energy-independent may be some restrictive.
In the future, with higher-statistics measurements of vector meson photoproduction from nucleus near-threshold,
we will further consider some corrections (such as incident-photon energy dependence and final-state interactions).
These refinements will improve the precision of the extracted vector meson–nucleus scattering lengths
and provide deeper insights into the in-medium modifications of vector mesons, as well as the strength and mechanisms of hadron–nucleon interactions.
}

In addition, we systematically explored the dependence of the vector meson–nucleus scattering length $|\alpha_{VN}|$ on several physical quantities,
including the vector meson mass, the target atom mass, and the threshold energy $\sqrt{s_{\mathrm{thr}}} = (m_V + m_T)^2$.
Our results show an approximate exponential suppression of $|\alpha_{VN}|$ with increasing vector meson mass and target atom mass,
indicating that heavier mesons (e.g., $J/\psi$, $\psi(2S)$, and $\Upsilon$) and heavier nuclei ($^{4}$He) exhibit weaker couplings in vector meson–nucleus interactions.
Furthermore, by comparing $\phi$-meson scattering lengths across various targets, a clear hierarchy $|\alpha_{\phi N}| > |\alpha_{\phi d}| > |\alpha_{\phi ^4\mathrm{He}}|$ was established,
reinforcing the mass-dependent nature of the interaction.
We also examined empirical correlations between $|\alpha_{VA}|$ and the inverse threshold energy $\sqrt{s_{\mathrm{thr}}}$,
as well as a rescaled variable $\left( \frac{1}{\sqrt{s_{\mathrm{thr}}}} \right) \times \left( \frac{m_T}{m_V} \right)^2$.
Both dependencies reveal potential universal scaling behavior in meson–nucleus scattering processes, and were parametrized using phenomenological exponential functions.
These findings hint at deeper connections between meson-nucleus interaction dynamics and QCD-scale phenomena, such as the role of the trace anomaly in generating nucleon mass.

\textcolor{black}{The scaling laws for the vector meson-nucleon and vector meson-nucleus scattering lengths $|\alpha_{VA}|$,
  as depicted by the dashed lines in Fig.~\ref{fig:avN_1mv}, \ref{fig:aphiN_1mT}, \ref{fig:avN_1sqrtSthr}, \ref{fig:avN_1sqrtSthr_mN2_mV2}, are empirical trends,
  rather than firm theoretical derivations from QCD.}
To further test these empirical trends and deepen our understanding of vector meson–nuclear interactions, we propose future near-threshold photoproduction experiments
involving a variety of vector mesons on helium-4 and other nuclear targets.
A broader set of high-precision vector meson photoproduction data on various nuclear targets is essential, and Jefferson Lab provides a suitable platform for conducting such experiments.
Upcoming experimental facilities, such as the Electron-Ion Collider (EIC) in the United States~\cite{Accardi:2012qut} and the Electron-Ion Collider
in China (EicC)~\cite{Chen:2018wyz,Chen:2020ijn,Anderle:2021wcy}, will provide ideal platforms for such measurements by utilizing high-luminosity virtual photon beams near threshold.
These future studies will not only serve as rigorous tests of the VMD model but also offer new insights into understanding the properties of hadronic interactions.

\begin{acknowledgments}
This work is supported by the National Natural Science Foundation of China under the Grant NO. 12305127,
the International Partnership Program of the Chinese Academy of Sciences under the Grant NO. 016GJHZ2022054FN,
the Research Program of State Key Laboratory of Heavy Ion Science and Technology, Institute of Modern Physics, Chinese Academy of Sciences, under Grant NO. HIST2025CS08,
and National Key R$\&$D Program of China under the Grant NO. 2024YFE0109802.
\end{acknowledgments}

\bibliographystyle{apsrev4-1}
\bibliography{refs.bib}

\end{document}